\documentclass[conference,letterpaper]{IEEEtran}

\usepackage[utf8]{inputenc} 
\usepackage[T1]{fontenc}
\usepackage{url}
\usepackage{ifthen}
\usepackage{comment}
\usepackage{graphicx}
\usepackage{cite}
\usepackage{amssymb}
\usepackage[nolist]{acronym}
\usepackage[cmex10]{amsmath} 

\newtheorem{theorem}{Theorem}

\newtheorem{remark}{Remark}

\begin{acronym}
    \acro{GDI}{General Definition of Information}
    \acro{IRP}{Inverse Relationship Principle}
    \acro{AEP}{Asymptotic Equipartition Property}
    \acro{p.m.f.}{probability mass function}
    \acro{LoA}{Levels of Abstraction}
    \acro{i.i.d.}{independent and identically distributed}
    \acro{CSI}{channel state information}
    \acro{CSIT}{channel state information available at the transmitter}
    \acro{CSIR}{channel state information available at the receiver}
    \acro{DMC}{discrete memoryless channel}
\end{acronym}
\begin{document}
\title{Semantic Communication Through the Lens of Context-Dependent Channel Modeling} 


\author{%
  \IEEEauthorblockN{Javad Gholipour, Rafael F. Schaefer, and Gerhard P. Fettweis}
  \IEEEauthorblockA{
  Technische Universität Dresden, Dresden, Germany\\
  Email: javad.gholipour@tu-dresden.de, rafael.schaefer@tu-dresden.de, gerhard.fettweis@tu-dresden.de}
}

\maketitle
\begin{abstract}
Semantic communication has emerged as a promising paradigm for next-generation networks, yet several fundamental challenges remain unresolved. Building on the probabilistic model of semantic communication and leveraging the concept of context, this paper examines a specific subclass of semantic communication problems, where semantic noise originates solely from the semantic channel, assuming an ideal physical channel. To model this system, we introduce a virtual state-dependent channel, where the state—representing context—plays a crucial role in shaping communication. We further analyze the representational capability of the semantic encoder and explore various semantic communication scenarios in the presence of semantic noise, deriving capacity results for some cases and achievable rates for others.
\end{abstract}


\section{Introduction}
Studies on the concept of (semantic) information, particularly from a philosophical perspective, date back centuries. However, despite significant efforts, the topic remains largely underexplored. A major breakthrough occurred in 1948 \cite{Shannon}, when Shannon introduced a quantitative measure of information known as entropy and established fundamental limits on source compression and the reliable transmission of data over noisy channels. His work provided communication engineers with a mathematical foundation to quantify information and optimize their systems performances. Although many philosophers argue that Shannon’s theory is not a true theory of information—and it has been reported that Shannon himself regretted the widespread labeling of his work as "information theory"—many researchers continue to view his communication model as a rigorous framework that constrains any further theorization of the concept of information \cite{Sommaruga}.

One year after Shannon's groundbreaking work \cite{Shannon}, Weaver proposed classifying communication problems into three subproblems: technical, semantic, and effectiveness \cite{Weaver}. The technical problem addresses the question: "How accurately can the symbols of communication be transmitted?" The semantic problem concerns the precision with which the transmitted symbols convey their intended meaning. However, communication does not end with the transmission of meaning; it also influences the receiver's behavior. The effectiveness problem, therefore, focuses on how successfully the conveyed meaning impacts the informee’s actions. He also noted that Shannon's model is sufficiently general to accommodate the semantic and effectiveness levels by incorporating relevant components.

Inspired by Weaver \cite{Weaver}, various efforts have aimed to extend Shannon's theory to the semantic and effectiveness levels of communication. Despite this, a comprehensive theory comparable to Shannon’s remains elusive. One direction involves defining semantic information via entropy-like measures. Bar-Hillel and Carnap \cite{Carnap} proposed a logical probability-based metric, but it is limited to propositional logic and leads to contradictions in specific cases. Floridi addressed these issues by adding truthfulness \cite{Floridi:2}, yet his model also remains constrained and dependent on external truth references. Overall, most semantic information measures are tailored to narrow domains and lack general applicability \cite{Gunduz}.

In a parallel effort, Bao et al. extended Shannon’s framework to include semantic elements, modeling communication within propositional logic \cite{Bao}. They formulated semantic source and channel coding using Bar-Hillel and Carnap's measure \cite{Carnap}. Ma et al. \cite{Ma} defined semantic channel capacity by treating semantics as a subset of the message set and transmitted these over noisy channels. However, we contend that natural semantic communication begins with generating semantics, followed by expressing them as messages, and ends with recovering semantics from received messages.


In \cite{Gholipour}, a probabilistic model for semantic communication was proposed, inspired by Weaver’s extension approach \cite{Weaver} and grounded in the \ac{GDI} \cite{Floridi:1,Sommaruga}, incorporating its key components such as well-formed data and meaningfulness. Unlike the model proposed in \cite{Shao} relying solely on intuitions derived from human language, \cite{Gholipour} is fundamentally rooted in a rigorous philosophical foundation and is thus applicable to the broader concept of information and data beyond linguistic communication. This model builds on the fundamental concept of data as Constraining Affordances and its relationship to meaning, mediated by the \ac{LoA} \cite{Sommaruga}. Within this framework, context is defined as the sufficient amount of information that ensures each message (data) has only one meaning, though its interpretation may vary across different contexts \cite{Gholipour}. In \cite{Gholipour}, the semantic communication problem—recognized as inherently complex—was proposed to be systematically addressed by categorizing it into simpler subproblems. Additionally, an achievable rate for reliable semantic communication was derived for the subproblem, where semantic noise arises solely from the physical channel noise. It was also shown that Shannon's model \cite{Shannon} is an extreme case of the proposed semantic communication model, where each message corresponds to a single, unambiguous semantic meaning, and each semantic is expressed by a unique message, establishing a strict one-to-one relationship between semantics and messages. Furthermore, it was proven that the semantic communication approach can enhance the rate of transmissible messages compared to Shannon's capacity, the semantic rate equals to Shannon's capacity though. In this paper, we take a step forward by analyzing the next subproblem, specifically the case where semantic noise arises solely from the semantic channel noise, with no physical channel noise present.

Noise is a key factor that limits (semantic) communication, making its study vital for theoretical advancement. While (physical channel) noise in traditional communication systems is well-understood, the concept of semantic noise has yet to be more explored. Shi et al. \cite{Shi} argued that semantic noise can arise at different stages of communication, including encoding, transmission, and decoding. Among others, Qin et al. \cite{Qin} defined semantic noise as disturbances that hinder the accurate interpretation of messages, while Hu et al. \cite{Hu} highlighted misalignments between the intended and received semantic symbols. Based on \cite{Shi}, in the semantic encoding stage, the semantic noise can lead to ambiguity in interpreting the message expressing a semantic, reflecting the encoder's capability to express the semantic unambiguously. During the data transmission stage, semantic noise is introduced by signal distortion caused by the physical channel noise. For example, physical channel noise can distort a word during transmission, e.g., changing "light" to "right", altering its intended meaning. In the decoding stage, semantic noise, stemming from semantic channel noise, is caused by discrepancies in knowledge between the sender and receiver, characterized as mismatched codebooks between them \cite{Gholipour}, leading to misinterpretations even when the transmitted signal is received without errors. In other words, semantic channel noise arises from context-dependent variations in meaning, such as the word "date," which can refer to a specific day on the calendar, a social meeting, or a type of fruit, depending on the context.

In this paper, building on the probabilistic model of semantic communication introduced in \cite{Gholipour} and leveraging the concept of context, we focus on analyzing the semantic communication system where semantic noise arises solely from the semantic channel, assuming a noiseless physical channel. As we discuss in detail in the following sections, we propose modeling such a system using a virtual state-dependent channel, where the state—representing the context—plays a crucial role in shaping the communication process. 

State-dependent channels with random states serve as a fundamental model in information theory. The capacity of these channels depends on whether \ac{CSI} is available at the transmitter, receiver, both, or neither, as well as whether it is known causally or non-causally. The study of state-dependent channels originated with Shannon \cite{Shannon2}, who introduced a model where \ac{CSI} is causally available at the transmitter and derived its capacity. Later, Gelfand and Pinsker \cite{Gelfand} extended this framework to the case where \ac{CSI} is available non-causally at the transmitter, demonstrating that the transmitter can mitigate state effects using a binning strategy. Motivated by the problem of memory with defective cells, Heegard and El Gamal \cite{Heegard} analyzed the capacity of a state-dependent \ac{DMC} where the defect information (acting as \ac{CSI}) is fully available at either the transmitter or the receiver. Further, Cover \cite{Cover2} investigated state-dependent channels where \ac{CSI} is only partially available at the transmitter and receiver. 

\textit{Notation}: we use capital letters to show random variables (e.g., $X$); lower case letters for realization of random variables (e.g., $x$) and $p(x)$ to show the \ac{p.m.f.} of random variable $X$ on set $\mathcal{X}$. We also define $X^n$ as an $n$-sequence. Also the set of $\epsilon$-strongly joint-typical of $n$-sequences $X^n$ and $Y^n$ with \ac{p.m.f.} $p(x,y)$, is shown by $T_\epsilon^n(X,Y)$. 
\section{Main Results}
In this section, we investigate the semantic communication problem in which semantic noise arises solely from the semantic channel, assuming a noiseless physical channel. First of all, we shortly review the probabilistic model for semantic communication \cite{Gholipour}, then introduce a semantic channel model inspired by the state-dependent channel, where the state is replaced by context. Next, we analyze the representational capability of the encoder, and finally, we examine various cases of semantic communication over the proposed model.

\vspace{-0.05em}
\subsection{Preliminaries}
\begin{figure}
\centering
\includegraphics[width=0.45\textwidth]{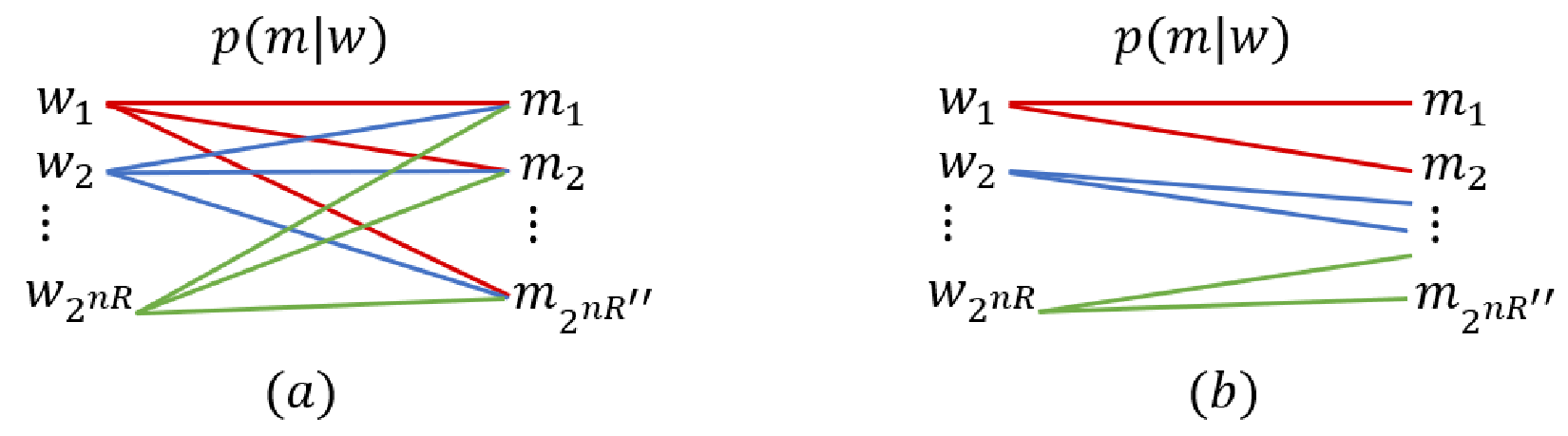}
\caption{Interrelationship between semantics $W$ and the messages $M$ \cite{Gholipour}, (a): in general, and (b): for a given context $Q=q$.}
\label{Figure1}
\vspace{-\baselineskip}
\end{figure}

According to \cite{Gholipour}, leveraging the \ac{GDI} and the concept of data as Constraining Affordances, the relationship between semantics and messages (data) can be probabilistically modeled as a virtual noisy channel between semantics and the messages. As illustrated in Figure \ref{Figure1}(a), according to the concept of data as Constraining Affordances \cite{Sommaruga}, each message $m$ can be interpreted as different semantics $w$ with the conditional probability $p(w|m)$, while each semantic $w$ can, in turn, be expressed through multiple messages $m$ with the conditional probability $p(m|w)$. Consequently, the interpretation of a message inherently involves ambiguity. The crucial factors mediating this relationship are the \ac{LoA} \cite{Sommaruga} and the context \cite{Gholipour}. Based on this idea, and motivated by the fact that right after generating a message by a semantic source, its meaning is at least clear for itself, \cite{Gholipour} proposed defining context as the sufficient amount of information needed to eliminate ambiguity in the interpretation of a message. Intuitively, in human communication, context encompasses a complex combination of factors such as time, location, the psychological state of the informer, etc. These elements collectively ensure that a given message carries a single, unambiguous meaning. Mathematically, the relationship between semantics, messages and context is defined as follows \cite{Gholipour}:
\begin{align}
    &H(W|M)\geq 0,\\
    &H(W|M,Q)=0,
\end{align}
where, the random variables $W$, $M$, and $Q$ represent the semantic, message, and the context, respectively.

As illustrated in Figure \ref{Figure1}(b), context plays a crucial role in eliminating the ambiguity by partitioning the set of messages into disjoint subsets, each of them associated with one semantic. While each semantic can be expressed by multiple messages, context ensures that each message is interpreted as a single, unambiguous semantic. A change in context alters this partitioning. For example, under context $q_1$, the message $m_1$ may be interpreted as semantic $w_1$, whereas under a different context $q_2$, the same message $m_1$ could be interpreted as another semantic $w_2$, reflecting the change in the partitioning way of the message set.
\subsection{Semantic Channel Model}
\begin{figure}
\centering
\includegraphics[width=0.45\textwidth]{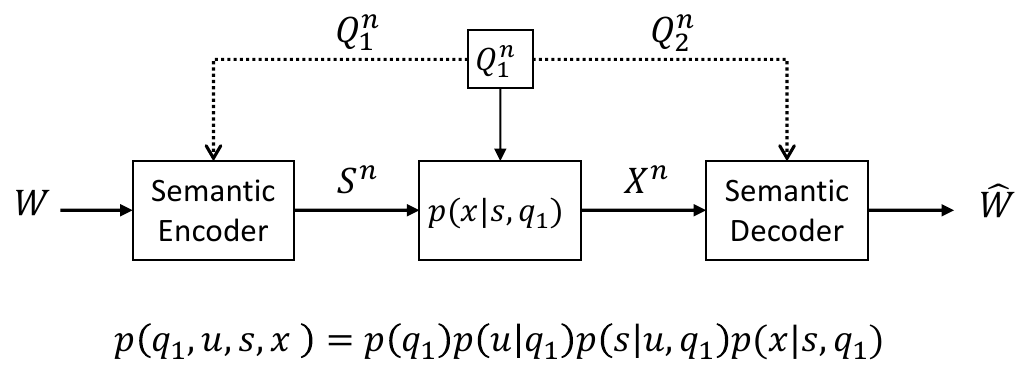}
\caption{Semantic channel modeled as a context-dependent channel.}
\label{Fig1}
\vspace{-\baselineskip}
\end{figure}

As defined in \cite{Shi} and discussed in the previous section, semantic noise refers to factors that introduce ambiguity in message interpretation and can occur at various stages of communication. 
In classical communication, channel coding is used to mitigate physical channel noise, which motivates us to explore a similar approach for combating semantic channel noise. In this paper, based on the probabilistic model describing the interrelationship between semantics and messages and motivated by the role of context as a crucial mediating factor \cite{Gholipour}, we propose modeling a semantic communication problem where semantic noise originates exclusively from the semantic channel (assuming a noiseless physical channel) using the state-dependent channel model, where the state is replaced by the context, as illustrated in Figure \ref{Fig1}. In this model, $W$ represents the semantic, and $\hat{W}$ is the reconstructed semantic. The semantic encoder takes the semantic $W$ and the available context $Q_1^n$ as the inputs, producing a semantic codeword $S^n$. Then, based on $S^n$ and $Q_1^n$, it generates a message codeword $X^n$ associated with a message $M$, expressing the intended semantic $W$, according to the conditional \ac{p.m.f.} $p(x|s,q_1)$. Since the physical channel is assumed noiseless, $X^n$ is received directly by the receiver, which, using its available context $Q_2^n$, reconstructs the semantic $\hat{W}$. Depending on the relationship between $Q_1^n$ and $Q_2^n$, various types of semantic communication problems can be modeled, in which semantic noise arises solely from the semantic channel noise. It is worth noting that the model illustrated in Figure \ref{Fig1} can also represent the semantic sender itself, in the special case where $Q_2^n=Q_1^n$, and the semantic sender knows its generated message codeword $X^n$.

\subsection{Representationally Capable Semantic Sender}

In \cite{Gholipour}, the impact of semantic noise at the transmitter, which affects its representational capability, was examined. To simplify notation, the context was omitted from the mathematical formulations. In this subsection, we revisit this problem shortly through the lens of a state-dependent channel model depicted in Figure \ref{Fig1}. Specifically, the semantic sender can be modeled as a state-dependent channel, where the channel state information—corresponding to the context—is assumed to be non-causally available at both the transmitter and the receiver, e.g., $Q_2^n=Q_1^n$. Consistent with the results in \cite{Gholipour}, we establish that a semantic sender can express itself for itself without ambiguity and is representationally capable if and only if its semantic rate does not exceed its representational capacity.

\begin{theorem}\label{Rep}
The semantic sender illustrated in Figure \ref{Fig1}, is representationally capable if and only if the semantic information rate does not exceed its representational capacity:
    \begin{align}
        C_{RC} = \max_{p(s\vert q_1)}\, I(S;X\vert Q_1).
    \end{align}
\end{theorem}

The proof follows a similar approach to that of the capacity analysis for state-dependent \ac{DMC}, as presented in \cite{Cover2}. In the following, we provide an outline of the achievability part of the proof:
\begin{itemize}
    \item Generate $2^{n\Tilde{R}}$ \ac{i.i.d.} auxiliary sequences $U^n$ according to the \ac{p.m.f.} $p(u)$.
    \item Partition the set of $2^{n\Tilde{R}}$ sequences $U^n$ into $2^{nR}$ equal-sized bins. Therefore, each bin contains $2^{n(\Tilde{R}-R})$ sequences $U^n$.
    \item Associate each semantic $w$ to a bin.
    \item For the semantic $w$ and the context $q_1^n$, look for a codeword $u^n$ inside the associated bin with $w$, such that:
    \begin{align}
        (u^n,q_1^n)\in T_\epsilon^n(U,Q_1).
    \end{align}
    \item The probability of not finding such a sequence $u^n$ inside the bin asymptotically tends to zero, if:
    \begin{align}\label{Tx}
        \Tilde{R}-R>I(U;Q_1).
    \end{align}
    \item Assume $p(s|u,q_1)$ is a deterministic function $\textit{f}$. Therefore, given $u^n$ and $q_1^n$, the semantic encoder generates the semantic codeword $s^n$ according to $s=\textit{f}(u,q_1)$.
    \item Given the semantic codeword $s^n$ and the context $q_1^n$, the semantic source express itself via a message codeword $X^n$ associated with a message $M$, according to the conditional \ac{p.m.f.} $p(x|s,q_1)$.
    \item Given the context $q_2^n=q_1^n$ and the received message codeword $x^n$, the semantic decoder looks for a unique $u^n$, such that:
    \begin{align}
        (u^n,x^n,,q_1^n)\in T_\epsilon^n(U,X,Q_1).
    \end{align}
    \item The probability of not finding such a unique sequence $u^n$ asymptotically tends to zero, if:
    \begin{align}\label{Rx}
        \Tilde{R}<I(U;X,Q_1).
    \end{align}
    \item Combining (\ref{Tx}) and (\ref{Rx}), the semantic source is unambiguously expressible, and therefore is representational capable, if: 
    \begin{align}
        R<I(U;X|Q_1).
    \end{align}
    \item Due to the Markov chain $U\rightarrow S\rightarrow X$ conditioned on $q_1$:
    \begin{align}
        R<I(U;X|Q_1)\leq I(S;X|Q_1).
    \end{align}
\end{itemize}

\subsection{Semantic Communication over a Semantic Channel}
As mentioned earlier, the model illustrated in Figure \ref{Fig1} can represent not only the semantic sender for analyzing its representational capability, but also a semantic communication scenario where semantic noise arises solely from the discrepancy between the transmitter's and receiver's background knowledge. This discrepancy is modeled through the availability of context information, while the absence of physical channel noise ensures that $X^n$ is directly received without distortion. In what follows, we analyze three possible scenarios: (1) the sender and receiver have similar background knowledge, (2) the receiver’s background knowledge is a subset of the sender’s knowledge, (3) the sender’s background knowledge is a subset of the receiver’s knowledge, and (4) the Sender and receiver share partial background Knowledge.

\subsubsection{Sender and Receiver with Similar Background Knowledge}
Consider the scenario where the sender and the receiver have the same context information.

\begin{theorem}\label{c10}
    The capacity of the semantic channel depicted in Figure \ref{Fig1}, which accounts for semantic noise while assuming a noiseless physical channel, in the scenario where the sender and the receiver have the same context information, e.g., $Q_2^n=Q_1^n$, is given by:
    \begin{align}
        C = \max_{p(s\vert q_1)}\, I(S;X\vert Q_1).
    \end{align}
\end{theorem}

The proof follows a similar approach to that of the capacity analysis for state-dependent \ac{DMC}, as presented in \cite{Cover2}. The outline of the achievability proof follows exactly the same structure as that of Theorem \ref{Rep}.

\begin{figure}
\centering
\includegraphics[width=0.45\textwidth]{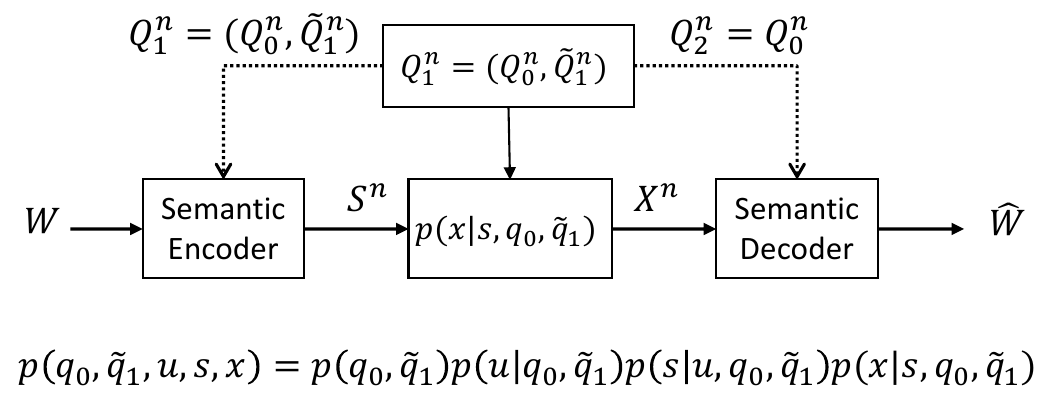}
\caption{Semantic channel modeled as a context-dependent channel, where the semantic receiver's background knowledge is a subset of the semantic sender's knowledge.}
\label{FigB}
\vspace{-\baselineskip}
\end{figure}

\subsubsection{Receiver’s Background Knowledge as a Subset of the sender’s Knowledge}
Consider the scenario where the receiver's context information is already known to the sender, as depicted in Figure \ref{FigB}—analogous to a mother communicating with her little child, where the mother is aware of the child's limited contextual understanding and adjusts her communication accordingly.

\begin{theorem}\label{c11}
    The capacity of the semantic channel depicted in Figure \ref{FigB}, which accounts for semantic noise while assuming a noiseless physical channel, in the scenario where the receiver's context information is already known to the transmitter, e.g., $Q_1^n=(Q_0^n,\Tilde{Q}_1^n)$ and $Q_2^n=Q_0^n$, is given by:
    \begin{align}
        C = I(U;X\vert Q_2)-I(U;\Tilde{Q}_1|Q_0).
    \end{align}
\end{theorem}

The proof follows a similar approach to that of the capacity analysis for state-dependent \ac{DMC}, as presented in \cite{Cover2}. In the following, we provide an outline of the achievability part of the proof:

\begin{itemize}
    \item Generate $2^{n\Tilde{R}}$ \ac{i.i.d.} auxiliary sequences $U^n$ according to the \ac{p.m.f.} $p(u)$.
    \item Partition the set of $2^{n\Tilde{R}}$ sequences $U^n$ into $2^{nR}$ equal-sized bins. Therefore, each bin contains $2^{n(\Tilde{R}-R})$ sequences $U^n$.
    \item Associate each semantic $w$ to a bin.
    \item For the semantic $w$ and the context $q_1^n=(q_0^n,\Tilde{q}_1^n)$, look for a codeword $u^n$ inside the associated bin with $w$, such that:
    \begin{align}
        (u^n,q_0^n,\Tilde{q}_1^n)\in T_\epsilon^n(U,Q_0,\Tilde{Q}_1).
    \end{align}
    \item The probability of not finding such a sequence $u^n$ inside the bin asymptotically tends to zero, if:
    \begin{align}\label{Tx1}
        \Tilde{R}-R>I(U;Q_0,\Tilde{Q}_1).
    \end{align}
    \item Assume $p(s|u,q_0,\Tilde{q}_1)$ is a deterministic function $\textit{f}$. Therefore, given $u^n$ and $(q_0^n,\Tilde{q}_1^n)$, the semantic encoder generates the semantic codeword $s^n$ according to $s=\textit{f}(u,q_0,\Tilde{q}_1)$.
    \item Given the semantic codeword $s^n$ and the context $(q_0^n,\Tilde{q}_1^n)$, the semantic source express itself via a message codeword $X^n$ associated with a message $M$, according to the conditional \ac{p.m.f.} $p(x|s,q_0,\Tilde{q}_1)$.
    \item Given the context $q_2^n=q_0^n$ and the received message codeword $x^n$, the semantic decoder looks for a unique $u^n$, such that:
    \begin{align}
        (u^n,x^n,,q_0^n)\in T_\epsilon^n(U,X,Q_0).
    \end{align}
    \item The probability of not finding such a unique sequence $u^n$ asymptotically tends to zero, if:
    \begin{align}\label{Rx1}
        \Tilde{R}<I(U;X,Q_0).
    \end{align}
    \item Combining (\ref{Tx1}) and (\ref{Rx1}), the semantic source is unambiguously expressible, if: 
    \begin{align}
        R<I(U;X|Q_0)-I(U;\Tilde{Q}_1|Q_0),
    \end{align}
    which, because $Q_2=Q_0$, simplifies to:
    \begin{align}
        R<I(U;X|Q_2)-I(U;\Tilde{Q}_1|Q_0).
    \end{align}
    \item Due to the Markov chain $U\rightarrow S\rightarrow X$ conditioned on $(q_0,\Tilde{q}_1)$:
    \begin{align}
        R<I(S;X|Q_2)-I(U;\Tilde{Q}_1|Q_0)
    \end{align}
\end{itemize}

\begin{figure}
\centering
\includegraphics[width=0.45\textwidth]{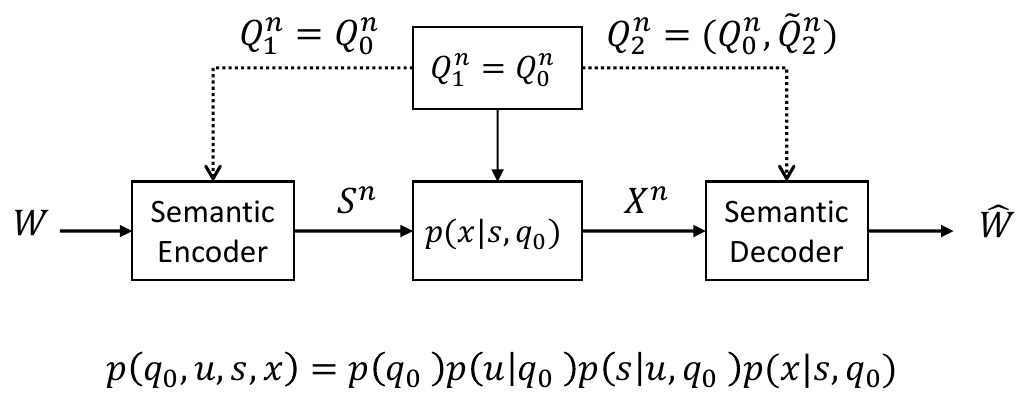}
\caption{Semantic channel modeled as a context-dependent channel, where the semantic sender’s background knowledge is a subset of the semantic receiver’s knowledge.}
\label{FigC}
\vspace{-\baselineskip}
\end{figure}

\subsubsection{Sender’s Background Knowledge as a Subset of the Receiver’s Knowledge}
Consider the scenario where the sender's context information is already known to the receiver, as depicted in Figure \ref{FigC}—analogous to a little child communicating with the mother, where the mother is aware of the child's limited contextual understanding and adjusts her interpretation accordingly, enabling her to comprehend the intended meaning even though the child’s speech may appear nonsensical to others.

\begin{theorem}\label{c12}
    For the semantic channel depicted in Figure \ref{FigC}, which accounts for semantic noise while assuming a noiseless physical channel, in the scenario where the sender's context information is already known to the receiver, e.g., $Q_1^n=Q_0^n$ and $Q_2^n=(Q_0^n,\Tilde{Q}_2^n)$, the following rate is achievable:
    \begin{align}
        R < I(U;X\vert Q_2)+I(U;\Tilde{Q}_2|Q_0).
    \end{align}
\end{theorem}

The proof is inspired by the similar approach to the capacity analysis of state-dependent \ac{DMC}, as presented in \cite{Cover2}.  In the following, we provide an outline of the proof:

\begin{itemize}
    \item Generate $2^{n\Tilde{R}}$ \ac{i.i.d.} auxiliary sequences $U^n$ according to the \ac{p.m.f.} $p(u)$.
    \item Partition the set of $2^{n\Tilde{R}}$ sequences $U^n$ into $2^{nR}$ equal-sized bins. Therefore, each bin contains $2^{n(\Tilde{R}-R})$ sequences $U^n$.
    \item Associate each semantic $w$ to a bin.
    \item For the semantic $w$ and the context $q_1^n=q_0^n$, look for a codeword $u^n$ inside the associated bin with $w$, such that:
    \begin{align}
        (u^n,q_0^n)\in T_\epsilon^n(U,Q_0).
    \end{align}
    \item The probability of not finding such a sequence $u^n$ inside the bin asymptotically tends to zero, if:
    \begin{align}\label{Tx2}
        \Tilde{R}-R>I(U;Q_0).
    \end{align}
    \item Assume $p(s|u,q_0)$ is a deterministic function $\textit{f}$. Therefore, given $u^n$ and $q_0^n$, the semantic encoder generates the semantic codeword $s^n$ according to $s=\textit{f}(u,q_0)$.
    \item Given the semantic codeword $s^n$ and the context $q_0^n$, the semantic source express itself via a message codeword $X^n$ associated with a message $M$, according to the conditional \ac{p.m.f.} $p(x|s,q_0)$.
    \item Given the context $q_2^n=(q_0^n,\Tilde{q}_2^n)$ and the received message codeword $x^n$, the semantic decoder looks for a unique $u^n$, such that:
    \begin{align}
        (u^n,x^n,,q_0^n,\Tilde{q}_2^n)\in T_\epsilon^n(U,X,Q_0,\Tilde{Q}_2).
    \end{align}
    \item The probability of not finding such a unique sequence $u^n$ asymptotically tends to zero, if:
    \begin{align}\label{Rx2}
        \Tilde{R}<I(U;X,Q_0,\Tilde{Q}_2).
    \end{align}
    \item Combining (\ref{Tx2}) and (\ref{Rx2}), the semantic source is unambiguously expressible, if: 
    \begin{align}
        R<I(U;X|Q_0,\Tilde{Q}_2)+I(U;\Tilde{Q}_2|Q_0),
    \end{align}
    which, because $Q_2=(Q_0,\Tilde{Q}_2)$, simplifies to:
    \begin{align}
        R<I(U;X|Q_2)+I(U;\Tilde{Q}_2|Q_0).
    \end{align}
\end{itemize}

\begin{figure}
\centering
\includegraphics[width=0.45\textwidth]{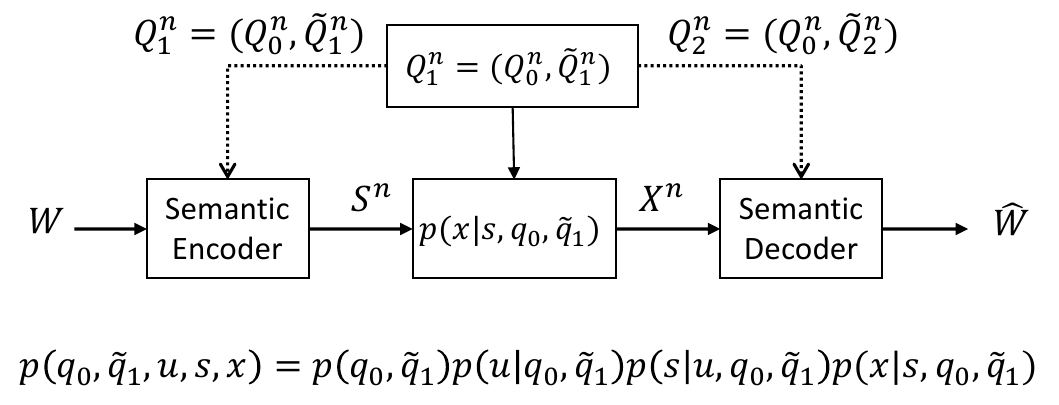}
\caption{Semantic channel modeled as a context-dependent channel, where the semantic sender and receiver share partial background knowledge.}
\label{FigD}
\vspace{-\baselineskip}
\end{figure}

\subsubsection{Sender and Receiver Share Partial Background Knowledge}
Consider the general scenario where the sender and receiver share partial context information, as depicted in Figure \ref{FigD}.

\begin{theorem}\label{c13}
    For the semantic channel depicted in Figure \ref{FigD}, which accounts for semantic noise while assuming a noiseless physical channel, in the general scenario where the sender and receiver share partial context information $Q_0^n$, which means that, $Q_1^n=(Q_0^n,\Tilde{Q}_1^n)$ and $Q_2^n=(Q_0^n,\Tilde{Q}_2^n)$, the following rate is achievable:
    \begin{align}
        R < I(U;X\vert Q_2)+I(U;\Tilde{Q}_2|Q_0)-I(U;\Tilde{Q}_1|Q_0).
    \end{align}
\end{theorem}

The proof is inspired by the similar approach to the capacity analysis of state-dependent \ac{DMC}, as presented in \cite{Cover2}.  In the following, we provide an outline of the proof:

\begin{itemize}
    \item Generate $2^{n\Tilde{R}}$ \ac{i.i.d.} auxiliary sequences $U^n$ according to the \ac{p.m.f.} $p(u)$.
    \item Partition the set of $2^{n\Tilde{R}}$ sequences $U^n$ into $2^{nR}$ equal-sized bins. Therefore, each bin contains $2^{n(\Tilde{R}-R})$ sequences $U^n$.
    \item Associate each semantic $w$ to a bin.
    \item For the semantic $w$ and the context $q_1^n=(q_0^n,\Tilde{q}_1^n)$, look for a codeword $u^n$ inside the associated bin with $w$, such that:
    \begin{align}
        (u^n,q_0^n,\Tilde{q}_1^n)\in T_\epsilon^n(U,Q_0,\Tilde{Q}_1).
    \end{align}
    \item The probability of not finding such a sequence $u^n$ inside the bin asymptotically tends to zero, if:
    \begin{align}\label{Tx3}
        \Tilde{R}-R>I(U;Q_0,\Tilde{Q}_1).
    \end{align}
    \item Assume $p(s|u,q_0,\Tilde{q}_1)$ is a deterministic function $\textit{f}$. Therefore, given $u^n$ and $(q_0^n,\Tilde{q}_1^n)$, the semantic encoder generates the semantic codeword $s^n$ according to $s=\textit{f}(u,q_0,\Tilde{q}_1)$.
    \item Given the semantic codeword $s^n$ and the context $(q_0^n,\Tilde{q}_1^n)$, the semantic source express itself via a message codeword $X^n$ associated with a message $M$, according to the conditional \ac{p.m.f.} $p(x|s,q_0,\Tilde{q}_1)$.
    \item Given the context $q_2^n=(q_0^n,\Tilde{q}_2^n)$ and the received message codeword $x^n$, the semantic decoder looks for a unique $u^n$, such that:
    \begin{align}
        (u^n,x^n,,q_0^n,\Tilde{q}_2^n)\in T_\epsilon^n(U,X,Q_0,\Tilde{Q}_2).
    \end{align}
    \item The probability of not finding such a unique sequence $u^n$ asymptotically tends to zero, if:
    \begin{align}\label{Rx3}
        \Tilde{R}<I(U;X,Q_0,\Tilde{Q}_2).
    \end{align}
    \item Combining (\ref{Tx3}) and (\ref{Rx3}), the semantic source is unambiguously expressible, if: 
    \begin{align}
        R<I(U;X|Q_0,\Tilde{Q}_2)+I(U;\Tilde{Q}_2|Q_0)-I(U;\Tilde{Q}_1|Q_0),
    \end{align}
    which, because $Q_2=(Q_0,\Tilde{Q}_2)$, simplifies to:
    \begin{align}
        R<I(U;X|Q_2)+I(U;\Tilde{Q}_2|Q_0)-I(U;\Tilde{Q}_1|Q_0).
    \end{align}
\end{itemize}

In summary, for various configurations of background knowledge, the following achievable rates are derived:

\begin{enumerate}
    \item Sender and receiver have similar knowledge:
    \begin{align}\label{sum1}
        R<I(U;X|Q_2),
    \end{align}
    \item When the receiver's knowledge is a subset of the sender's, the sender needs to adjust the communication by transmitting at a lower rate compared to the case in (\ref{sum1}), in order to remain understandable to the receiver.
    \begin{align}\label{sum2}
        R<I(U;X\vert Q_2)-I(U;\Tilde{Q}_1|Q_0).
    \end{align}
    \item When the sender's knowledge is a subset of the receiver's, the receiver can adapt the interpretation accordingly, allowing the sender to communicate at a higher rate compared to the case in (\ref{sum1}):
    \begin{align}\label{sum3}
        R<I(U;X\vert Q_2)+I(U;\Tilde{Q}_2|Q_0).
    \end{align}
    \item When the sender and receiver share partial knowledge, the achievable communication rate—compared to the case in (\ref{sum1})—can be either lower or higher, depending on the nature of their shared and private knowledge.
    \begin{equation}\label{sum4}
        R<I(U;X\vert Q_2)+I(U;\Tilde{Q}_2|Q_0)-I(U;\Tilde{Q}_1|Q_0).
    \end{equation}
\end{enumerate}

\section{Conclusion}
This paper investigated the evolving paradigm of semantic communication. In this paper, we explored a specific subclass of semantic communication problems, where semantic noise arises solely from the semantic channel, assuming a noiseless physical channel. Building on the probabilistic model of semantic communication and leveraging the role of context, we proposed a virtual state-dependent channel model to capture the impact of contextual information on communication. Through this model, we analyzed the representational capability of the semantic encoder and examined different semantic communication scenarios affected by semantic noise. Finally, we derived the capacity results for these scenarios, providing fundamental insights into reliable semantic communication in the absence of physical channel noise. These findings enhance our understanding of semantic communication and provide a foundation for future research aimed at unlocking its full potential for next-generation communication networks.

\end{document}